\documentstyle[aps,prl,multicol,epsf]{revtex}
\begin{document}

\def\datestamp{May 2, 2003}

\def\Top#1{\vskip #1\begin{picture}(290,80)(80,500)\thinlines
\put(65,500){\line(1,0){255}}\put(320,500){\line(0,1){5}}\end{picture}}
\def\Bottom#1{\vskip #1\begin{picture}(290,80)(80,500)\thinlines
\put(330,500){\line(1,0){255}}\put(330,500){\line(0,-1){5}}\end{picture}}
\def\top{\Top{-2.8cm}}
\def\bottom{\Bottom{-2.7cm}}

\title{Staggered-flux normal state in the weakly doped $t$-$J$ model}
\author{Dmitri~A.~Ivanov$^1$, Patrick~A.~Lee$^2$}
\address{
$^1$ Paul Scherrer Institute, CH-5232 Villigen PSI, Switzerland\\
$^2$ Department of Physics, M.I.T., Cambridge, MA 02139, USA}
\date\datestamp
\maketitle

\begin{abstract}
A normal (non-superconducting) ground state of the $t$-$J$ model
may be variationally approximated by a Gutzwiller-projected
wave function. Within this approximation, at small hole doping
near half-filling, the normal state favors staggered-flux ordering.
Such a staggered-flux state may occur in vortex cores of underdoped
high-temperature cuprate superconductors. From comparing the
energies of the staggered-flux state and of the superconducting state,
we numerically obtain the condensation energy. Extracting
the superfluid density directly from the projected
superconducting wave function, we can also estimate the coherence
length at zero temperature. 
\end{abstract}

\begin{multicols}{2}
Gutzwiller-projected (GP) wave functions are known to give
good variational energies for the $t$-$J$ model in the range
of parameters relevant for high-temperature cuprate 
superconductors\cite{Paramekanti,Yokoyama,Gros}. 
Not only they correctly predict the
$d$-wave symmetry of the superconducting pairing, but they also
successfully describe properties of the
superconducting state including doping dependence of the
order parameter, quasiparticle spectral weight, Drude weight,
and even antiferromagnetic instability at very
low doping\cite{Himeda}. 
Conceptually, use of GP wave functions for studying 
cuprate superconductors is tempting because of their 
resonating-valence-bond structure\cite{PWA}, which may be relevant
for such effects as topological order and spin-charge 
separation proposed for explaining unconventional properties
of underdoped cuprates\cite{Kivelson,Senthil}.

If we indeed assume that GP wave functions capture the
essential physics of underdoped cuprates, we may further
use such wave functions for describing not only their
superconducting, but also the {\it normal} state. 
While the ``pseudogap'' normal phase appears above the 
superconducting transition temperature and is not accessible
for the variational wave function approach, the
normal state also appears in vortex cores within the
superconducting phase. From the available experimental
evidence, the normal vortex cores are closer in their properties
to the pseudogap phase than to the 
conventional Fermi liquid\cite{Pan,Renner}.
Lee and Wen suggested that the normal state in the vortex core 
is a staggered-flux state\cite{Lee-vortex}. 
Such a state may be described
by GP variational wave functions, in a manner similar to the
superconducting state.

The main goal of this paper is to construct a normal
ground-state wave function of the $t$-$J$ model by projecting
the doped staggered-flux state and to compare the resulting
variational energy to that of the superconducting state.
To make the paper self-contained, we start with a brief
overview of the relations between projected staggered-flux
and superconducting wave functions. This part also explains
our motivation to use the staggered-flux wave function for
the normal state. The second part of the paper contains the
variational Monte Carlo results on the condensation energy
and their implications for the doping dependence 
of the coherence length.

At zero doping, the staggered-flux state and the $d$-wave
superconducting state yield the same variational wave function
upon Gutzwiller projection (projecting onto the no-double occupancy
states) due to the particle--hole symmetry\cite{Affleck,Zhang}. 
The resulting wave function describes a spin liquid with
the algebraic decay of spin correlations\cite{I-thesis}. 
This spin-liquid state
is not physically realized at zero doping because of the
antiferromagnetic (AF) instability leading to the AF ordering.
The use of GP wave functions for describing the ground state
of the $t$-$J$ model is based on the assumption that upon doping
this AF Mott insulator with holes, the AF instability disappears
and the spin-liquid behavior is restored.

The most used variational ansatz for the weakly doped $t$-$J$ model
is the {\it nearest-neighbor $d$-wave pairing} state involving
only nearest-neighbor hopping and nearest-neighbor $d$-wave pairing
on the square lattice\cite{Yokoyama,Gros}. 
For such a state, the equivalence of the
staggered-flux and the $d$-wave pairing states may be extended
to the case of non-zero doping, if the notion of Gutzwiller projection
is modified in a SU(2) invariant way (respecting the particle-hole
symmetry away from half filling)\cite{ILW}. The projected wave function has
algebraic decay of spin and current correlations. The algebraic
decay of correlation functions suggests that this wave function
may represent a critical point and not a stable phase. In our
further discussion we label this wave function as ``critical'' (CR).

It is known that the variational energy of the CR wave function may be
further lowered by adding a non-zero chemical potential before
projecting (in the pairing gauge)\cite{Yokoyama}.
In the mean-field theory, this chemical potential plays an important
role for stabilizing superconductivity\cite{Wen-Lee-SU2}. In the GP wave
function approach, the role of the chemical potential is less transparent,
here it serves only as an additional variational 
parameter.
It shifts the nodes in the spectrum from $(\pi/2,\pi/2)$ to
an incommensurate point along the diagonal of the Brillouin zone.
We conjecture that a non-zero chemical potential also cuts off
the algebraic behaviour of the correlation functions at a finite
correlation length, but this so far could not be convincingly proven
by numerical calculations limited to relatively small system sizes.
The GP wave function with the variationally optimized chemical
potential we further denote as ``superconducting'' (SC).

In fact, both the CR and SC wave functions are superconducting in
the sense that they break the U(1) electromagnetic gauge symmetry
in the thermodynamic limit (at a non-zero hole doping). While
this property is rather obvious for the SC state, it requires an
additional clarification for the CR state. As explained in 
Ref.~\onlinecite{ILW},
the CR state may be obtained by projecting an undoped staggered-flux
state wave function by means of the special ``SU(2)-invariant'' Gutzwiller
projection. Since the wave function before the projection is not
superconducting, one could doubt the superconducting nature of the
projected wave function. However, the SU(2)-invariant Gutzwiller
projection involves two species of slave bosons designed to convert both
empty and doubly-occupied sites into physical holes. In the
thermodynamic limit, these two species of bosons form two Bose
condensates. The relative phase between those condensates corresponds
to the broken U(1) electromagnetic gauge symmetry.

If one attempts to design a wave function of a superconducting vortex
with the use of Gutzwiller-projection, the broken U(1) symmetry in
the CR and SC wave functions comes into play: it is not possible to 
construct a smooth vortex core by a slow variation of SC or CR wave 
functions. A naive way to resolve this problem is to suppress the 
order parameter in the vortex core, as it happens in conventional
superconductors. However, as pointed out in Ref.~\onlinecite{Lee-vortex}, 
this may be not
the energetically cheapest vortex core. A more energetically favorable
vortex core could be constructed by modifying the CR wave function into
a non-superconducting one. For this purpose, we take the unprojected
undoped staggered-flux state used in Ref.~\onlinecite{ILW} 
for the SU(2)-invariant
Gutzwiller projection, and dope it until the number of fermions exactly
matches the required number of physical electrons (such a doping
opens Fermi pockets around the $(\pi/2,\pi/2)$ points of the Brillouin
zone). If we further apply the SU(2)-invariant Gutzwiller projection
to this doped staggered-flux wave function, only one of the two species
of bosons get involved (since the number of the fermions exactly matches
the required number of electrons, the doubly-occupied sites should not
be converted into holes), and the SU(2)-invariant Gutzwiller projection
in this case coincides with the usual one (prohibiting doubly-occupied
sites). The resulting state is obviously non-superconducting: it does
not break the electromagnetic U(1) symmetry. Instead it breaks the
time-reversal and translational symmetries, as it has static currents
circulating in the staggered-flux pattern. We further denote this GP
wave function as the ``staggered-flux'' (SF) state.

\end{multicols}
\top
%


\begin{figure}
\epsfxsize=0.7\hsize
\centerline{\epsffile{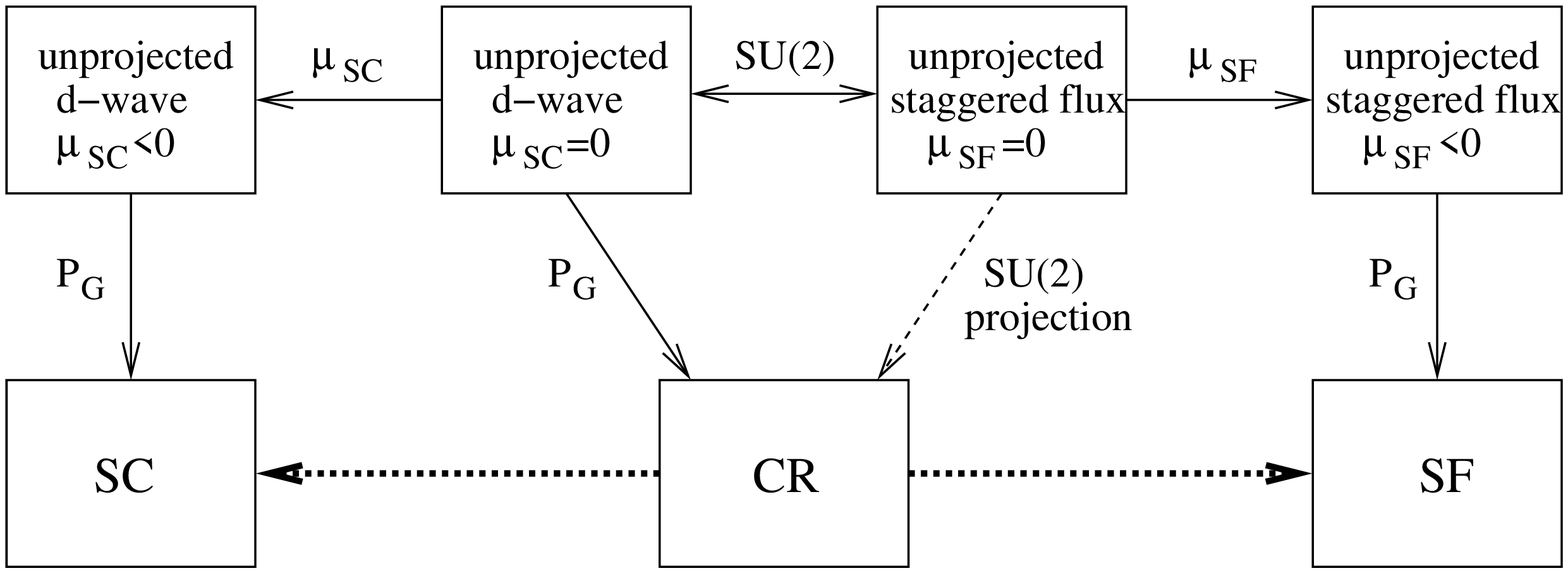}}
\bigskip
\caption{A schematic illustration of the construction of the GP wave
functions SC, CR, and SF. The top row of rectangles denotes unprojected
wave functions. The unprojected nearest-neighbor d-wave and
staggered flux states at $\mu_{\rm SC}=0$ and $\mu_{\rm SF}=0$ 
are related by a SU(2) gauge transformation in the particle-hole space.
Vertical solid arrows denote the Gutzwiller projection $P_G$, and the
dashed arrow is the SU(2)-invariant projection as defined 
in Ref.~\protect\onlinecite{ILW}. The dotted arrows connecting the CR state
to SC and SF states are drawn to illustrate that the two latter states
are continuous deformations of the CR state. }
\label{fig1}
\end{figure}


%
\bottom
\begin{multicols}{2}

We schematically summarize the relationship between those three types of the
GP wave functions (CR, SC, and SF) in Fig.~\ref{fig1}. From our construction
it follows that both the SC and SF states may be obtained as deformations
of the CR state (with the required deformation being small at small
doping). Therefore, at small doping, SC and SF states are close in energy,
and this makes the SF state a good candidate for the competing ground
state.

From comparing the energies of the SC state ($E_{\rm SC}$) and
of the SF state ($E_{\rm SF}$), we cad deduce the condensation energy
\begin{equation}
\varepsilon_c=E_{\rm SF}-E_{\rm SC}\ .
\end{equation}
The condensation energy is involved in the energy balance
determining the order parameter in non-uniform settings,
e.g.\ in superconducting vortices. The energy of a superconducting
state with a non-uniform phase of the order parameter may
be written in the Ginzburg-Landau form
\begin{equation}
E=\varepsilon_c + \rho_s (\nabla \varphi)^2 \ ,
\label{GL-energy}
\end{equation}
where $\varphi$ is the phase of the order parameter, and
$\rho_s$ is the superfluid stiffness (proportional to the
superfluid density). The size $\xi$ of the vortex core
may be estimated from minimizing the total energy consisting
of the two parts: the core energy $\pi\xi^2\varepsilon_c$
(up to a numerical prefactor of order one depending on the
specific shape of the order-parameter profile) and
the supercurrent energy $2\pi\rho_s\log \lambda/\xi$ (where $\lambda$
is the infrared cut-off). The resulting vortex size $\xi$
(which may also be called Ginzburg--Landau coherence length
at zero temperature) is
\begin{equation}
\xi=\sqrt{\rho_s/\varepsilon_c}\ .
\label{xi}
\end{equation}

The superfluid stiffness for strongly correlated systems was
discussed in detail in Ref.\onlinecite{Paramekanti-2}.
It is given by the sum of the diamagnetic term
(proportional to the kinetic energy in the ground state) and
of the paramagnetic term determined by the quasiparticle excitations.
For our superconducting state, at the mean-field level,
the low-lying quasiparticles have Dirac-like spectrum around 
the nodal points. We assume that the low-lying quasiparticles
preserve their mean-field structure, then the paramagnetic
contribution vanishes at zero temperature \cite{Lee-Wen-97}.
Thus $\rho_s$ is given by the diamagnetic term alone which,
in our notation, equals\cite{Paramekanti-2}
\begin{equation}
\rho_s = -{1\over16}\langle E_t \rangle \ ,
\label{rho-s}
\end{equation}
where $E_t$ is the hopping part of the $t$-$J$ Hamiltonian,
and the average is taken in the SC state.

Below we present our numerical results for $\varepsilon_c$
and $\rho_s$ (by
the variational Monte Carlo method) in the $t$-$J$
model with $t/J=3$.

We start with defining the variational parameters of the wave functions.
A GP wave function is constructed as
\begin{equation}
\Psi_{\rm GP}=P_G \Psi_0 \ ,
\end{equation}
where $P_G$ is the ``double'' projection: first, it projects out
components with doubly occupied sites (the usual Gutzwiller projection),
and second, it fixes the number of particles to the required value
(we shall work with finite systems where the required doping
will be enforced via projection). $\Psi_0$ is the ground-state
wave function of a BCS Hamiltonian:
\begin{equation}
 H = \sum_{ij} \left(-\chi_{ij} c^{\dagger}_{i\alpha}c_{j\alpha} 
 + \Delta_{ij} (c^{\dagger}_{i\uparrow} c^{\dagger}_{j\downarrow} -
 c^{\dagger}_{i\downarrow} c^{\dagger}_{j\uparrow}) + {\rm h.c.} \right).
\label{BCS-hamiltonian}
\end{equation}
$\chi_{ij}$ and $\Delta_{ij}$ are hopping and pairing amplitudes
variationally adjusted to minimize the expectation value of the
$t$-$J$ Hamiltonian
\begin{equation}
 H = P_G \left[
 \sum_{ij} \left(-t c^{\dagger}_{i\alpha}c_{j\alpha} 
 + J (\vec{S}_i \vec{S}_j - {1\over4} n_i n_j) \right) \right] P_G \ .
\label{t-J-hamiltonian}
\end{equation}
in the state $\Psi_{\rm GP}$.


\begin{figure}
\epsfxsize=0.7\hsize
\centerline{\epsffile{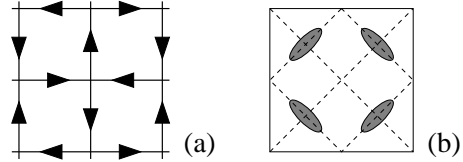}}
\bigskip
\caption{{\bf (a)} The vector potential in the staggered-flux state.
{\bf (b)} The Fermi pockets around $(\pi/2,\pi/2)$ points in the
staggered-flux state.}
\label{fig2}
\end{figure}


The CR state has $\chi_{ij}$ and $\Delta_{ij}$ non-zero only
on nearest-neighbor links: $\chi_{ij}=\chi$, $\Delta_{ij}=\pm\Delta$,
with $\pm$ for vertical and horisontal links respectively.
The SC state differs from the CR state only by the on-site term
$\chi_{ii}=-\mu_{\rm SC}$. The SF state has $\Delta_{ij}=0$,
$\chi_{ij}=e^{ia_{ij}}$, where $a_{ij}=\pm\Phi/4$ is the vector
potential defining the staggered flux pattern with the flux $\Phi$
(Fig.~\ref{fig2}a). The SF state also contains the chemical potential
$\chi_{ii}=-\mu_{\rm SF}$ which is fixed to provide 
the required hole density and is not a variational parameter
(unlike $\mu_{\rm SC}$ in the SC state).
At zero doping, all the three states coincide with 
$\mu_{\rm SC}=\mu_{\rm SF}=0$, $\Delta/\chi=\tan(\Phi/4)$.

\begin{figure}
\epsfxsize=1.0\hsize
\centerline{\epsffile{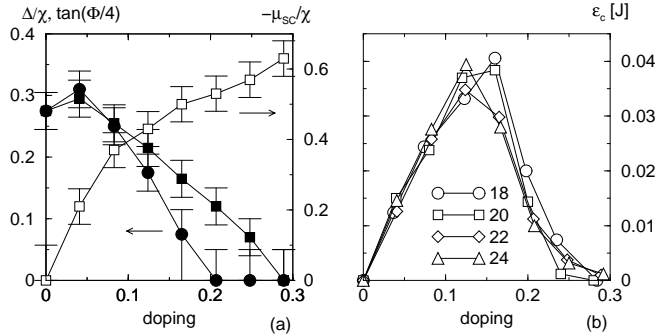}}
\smallskip
\caption{{\bf (a)} The gaps in the SF and SC states (solid circles
and squares, respectively, scale on the left side) and $\mu_{\rm SC}$
in the SC state (empty squares, scale on the right side) at different
hole dopings. The optimization is performed on the 22$\times$22 lattice with
boundary conditions periodic in one and antiperiodic in the other
direction. $t/J=3$.
{\bf (b)} The condensation energy $\varepsilon_c$ at different dopings
and for different system sizes ($N\times N$ lattice with $N=18,20,22,24$)
in the units of $J$, per lattice site.}
\label{fig3}
\end{figure}

The variational parameters are $\mu_{\rm SC}$ and $\Delta/\chi$
in the SC state, and $\Phi$ in the SF state. We determine these
parameters as a function of doping by minimizing the energy
on the 22$\times$22 lattice with the boundary conditions periodic in one
and antiperiodic in the other direction. The results are plotted
in Fig.~\ref{fig3}a. 
We find that while the
gap in the superconducting state closes at around 
30\% doping\cite{comment-gap}, the gap in the SF state
closes at a smaller doping (around 20\%).

We further use those variational parameters to determine 
the condensation energy $\varepsilon_c$. The finite-size effects
are very strong in the SF state, because the Fermi pockets 
(Fig.~\ref{fig2}b) are represented only by a small number 
of points in the momentum
space. To estimate the magnitude of the finite-size effects,
we plot $\varepsilon_c$ for different system sizes, but with the
same variational parameters, in Fig.~\ref{fig3}b. 
At small doping, $\varepsilon_c$
grows roughly linearly with doping. 
This linear doping dependence is not intuitive: the mean-field
theory would give $x^{3/2}$ dependence on the doping $x$,
from the energy of the Fermi pockets. Remarkably, the same linear
$x$ dependence was obtained by Lee and Nagaosa after
including the gauge-field fluctuations\cite{Lee-Nagaosa}.
As a result of this linear $x$-dependence, the core size
remains finite in the small-doping limit. As the doping
increases, the gaps in the SF and SC states decrease, 
which eventually leads to a decrease in the condensation 
energy $\varepsilon_c$.
When the gaps close, the SF and SC states again coincide
(with $\mu_{\rm SC}=\mu_{\rm SF}$), yielding $\varepsilon_c=0$.


The profile of $\varepsilon_c$ versus doping resembles the doping
dependence of $T_c$ in the cuprates. It seems reasonable to
interpret the regions of increasing and decreasing $\varepsilon_c$
as underdoped and overdoped regimes, respectively. With this interpretation,
our results indicate that in the uderdoped (and possibly also in the
weakly overdoped) regime the normal state inside the vortex core
has a staggered-flux order. This order disappears in the strongly
overdoped regime.


\begin{figure}
\epsfxsize=0.95\hsize
\centerline{\epsffile{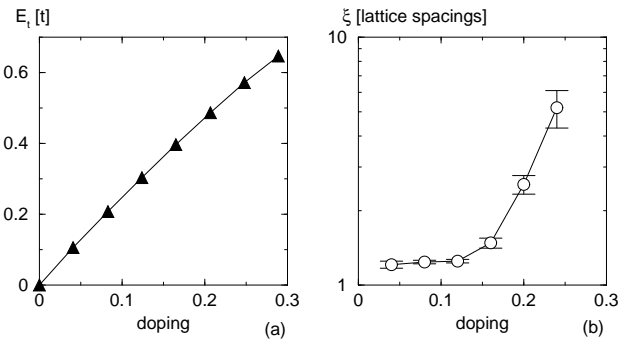}}
\smallskip
\caption{{\bf (a)} The hopping energy $E_t$ in the SC state
as a function of doping (in the units of $t$, per lattice site). 
The data shown are for the 22${\times}$22 lattice ($t/J=3$). 
The finite-size effects and the error bars are smaller than the
symbol size.
{\bf (b)} The coherence length $\xi$ as a function of the doping
for $t/J=3$. Note the logarithmic scale for $\xi$.}
\label{fig4}
\end{figure}


We further compute the superfluid stiffness $\rho_s$ using
Eq.(\ref{rho-s}).
In Fig.~\ref{fig4}a we plot the hopping energy $E_t$ in the SC state
as a function of doping (here we use the optimized values of $\Delta$
and $\mu_{\rm SC}$). The doping dependence of $\rho_s$ is
neary linear as expected\cite{Lee-Wen-97,comment-rho-s}.

Combining the results for $\varepsilon_c$ and for $\rho_s$, we
find the Ginzburg-Landau coherence length $\xi$ according to (\ref{xi}).
The results are shown in Fig.~\ref{fig4}b. 

Even though the staggered-flux core is relatively cheap in energy,
the resulting coherence length is very short in the underdoped
region. We find the coherence length of the order of one lattice
spacing, which is smaller than the experimental 
findings\cite{Pan,Renner,Wang}. 
Such a short coherence length must be considered a lower bound
only, because SF core of the size of one lattice spacing does 
not make any physical sense. When we approximated the core energy
by $\pi\xi^2\varepsilon_c$, we have used the bulk energy density
and ignored the cost of the boundary between the SF and SC, 
i.e.\ the energy of smoothly connecting the two states. This assumption
is correct only if the boundary is slowly varying and it surely breaks down
when the distance scale is about one lattice constant.

In our treatment we neglected the possible AF order which probably
plays a role at very low doping (below 0.1)\cite{Himeda}.
We expect that taking into account possible AF ordering both
in the normal and in the superconducting states lowers the energy of both
and only slightly modifies our results at the very low doping.

P.~A.~L.\ acknowledges support by NSF grant DMR-0201069.
Most of the numerical computations have been performed on the Beowulf
cluster Asgard at ETH Z\"urich. D.~I.\ thanks ETH Z\"urich for hospitality.

\end{multicols}

\begin{references}
\bibitem{comment-gap}
In Refs.\onlinecite{Yokoyama,Paramekanti}, the maximal doping 
admitting superconductivity is estimated about 36\% (at $t/J\sim 3$).
We find a slightly lower figure of 30\%. The discrepancy may
be due to a larger system size in our numerics. 
Ref.\onlinecite{Paramekanti} also uses next-nearest-neighbor hopping
in the $t$-$J$ model (not included in our work), which should lead
to quantitative differences at large dopings.
\bibitem{comment-rho-s}
The same linear doping dependence is reported in Ref.\onlinecite{Paramekanti}
(at large doping, our results slightly differ most probably because
of the next-nearest-neighbor hopping included in the model of
Ref.\onlinecite{Paramekanti}).
\end{references}
\end{document}